\documentclass[11pt]{article}

\usepackage{amsmath}
\usepackage{graphicx}
\usepackage{amsfonts}
\usepackage{amssymb}
\usepackage{epsfig}
\usepackage{color}
\usepackage{psfrag}
\usepackage{epstopdf}

\usepackage{caption}
\usepackage{subcaption}

\newcommand{\EQ}{\begin{equation}}
\newcommand{\EN}{\end{equation}}
\newcommand{\be}{\begin{equation}}
\newcommand{\ee}{\end{equation}}
\newcommand{\bea}{\begin{eqnarray}}
\newcommand{\eea}{\end{eqnarray}}

\DeclareMathOperator*{\SumInt}{%
\mathchoice%
  {\ooalign{$\displaystyle\sum$\cr\hidewidth$\displaystyle\int$\hidewidth\cr}}
  {\ooalign{\raisebox{.14\height}{\scalebox{.7}{$\textstyle\sum$}}\cr\hidewidth$\textstyle\int$\hidewidth\cr}}
  {\ooalign{\raisebox{.2\height}{\scalebox{.6}{$\scriptstyle\sum$}}\cr$\scriptstyle\int$\cr}}
  {\ooalign{\raisebox{.2\height}{\scalebox{.6}{$\scriptstyle\sum$}}\cr$\scriptstyle\int$\cr}}
}

\begin{document} \setcounter{page}{0}
\newpage
\setcounter{page}{0}
\renewcommand{\thefootnote}{\arabic{footnote}}
\newpage
\begin{titlepage}
\begin{flushright}
\end{flushright}
\vspace{0.5cm}
\begin{center}
{\large {\bf Quantum quenches with long range interactions}}\\
\vspace{1.8cm}
{\large Marianna Sorba$^{1}$, Nicol\`o Defenu$^{1,2}$ and Gesualdo Delfino$^{3,4}$}\\
\vspace{0.5cm}
{\em $^1$Institute for Theoretical Physics, ETH Z\"{u}rich, Wolfgang-Pauli-Str. 27, 8093 Z\"{u}rich, Switzerland}\\
{\em $^2$CNR-INO, Area Science Park, Basovizza, 34149 Trieste, Italy}\\
{\em $^3$SISSA -- Via Bonomea 265, 34136 Trieste, Italy}\\
{\em $^4$INFN sezione di Trieste, 34100 Trieste, Italy}\\
\end{center}
\vspace{1.2cm}

\renewcommand{\thefootnote}{\arabic{footnote}}
\setcounter{footnote}{0}

\begin{abstract}
\noindent

\end{abstract}
We extend the theory of quantum quenches to the case of $d$-dimensional homogeneous systems with long range interactions. This is achieved treating the long range interactions as switched on by the quench and performing the derivation within the basis of asymptotic states of the short range interacting pre-quench theory. In this way we analytically determine the post-quench state and the one-point functions of local observables such as the order parameter. One implication is that, as in the short range case, some oscillations induced by the quench remain undamped at large times under conditions specified by the theory. This explains, in particular, why such undamped oscillations have been numerically observed also in presence of long range interactions.
\end{titlepage}

\newpage
\tableofcontents

\section{Introduction}
The nonequilibrium quantum dynamics of isolated statistical systems is characterized by the fact that unitary time evolution occurs in a state which is a superposition of infinitely many eigenstates of the Hamiltonian. The simplest physical way to access such a nonequilibrium quantum state is to initially have the system in the ground state of its Hamiltonian $H_0$ and then suddenly change an interaction parameter to obtain the Hamiltonian $H$ which rules the time evolution thereafter \cite{BMcD}. In spite of its basic character, this procedure -- which has been called ``quantum quench" \cite{SPS,CC} -- proved to pose a nontrivial problem from the point of view of theoretical study. The way to determine the state generated by the quench was eventually found in \cite{quench} and allowed to reveal very interesting general features of the dynamics. In particular, it first appeared that oscillations\footnote{The study of quantum quenches was significantly boosted by the early experimental observation of long lived oscillations in \cite{KWW}.} of one-point functions (e.g. the order parameter) remain undamped at large times under conditions specified by the theory \cite{quench}. While initially derived perturbatively in the quench size\footnote{The interaction among the collective excitations of the system (quasiparticles) is arbitrarily strong.} \cite{quench,DV}, it was explained in \cite{oscill,oscillD} why this property actually holds nonperturbatively, and the general nonperturbative structure was exhibited in \cite{structure}. The undamped oscillations of \cite{quench} have been observed numerically \cite{BCH,KCTC,Lukin}, most remarkably in \cite{Jacopo}, where the time evolution was followed in simulations of the Ising spin chain reaching times several orders of magnitude larger than the timescales of the system. 

The remarkable predictive power of the theory introduced in \cite{quench} -- which covers any spatial dimensionality \cite{oscillD,structure}, inhomogeneous quenches \cite{oscill,oscillD}, and other initializations of the dynamics \cite{structure,q_int,excited} -- relies on the ability to cast the problem of quantum nonequilibrium in the formalism of asymptotic states of quantum field theory, a framework which allowed the general analysis of systems which in their microscopic (e.g. lattice) realization possess short range interactions. On the other hand, the numerical finding of \cite{Liu} that the nonequilibrim dynamics of a spin chain with {\it long range} interactions displayed features -- including undamped oscillations -- quite analogous to those of the short range case suggested that the domain of application of asymptotic state theory could be further extended. While this extension looks a priori awkward in view of the additional difficulties involved in the analytical study of long range interacting systems \cite{LR_review}, we show in this paper that it can be achieved in the case of quantum quenches. For this purpose we consider translation invariant systems in $d$ spatial dimensions with Hamiltonian of the form
\begin{equation}
\left\{
\begin{array}{l}
H_0\,,\hspace{.5cm}t<0\,,\\
\\
H=H_0+\lambda \int d\mathbf{x}\, \Psi(\mathbf{x})\,,\hspace{.5cm}t\geq 0\,,
\end{array}
\right.
\label{Hamiltonian}
\end{equation} 
where the pre-quench Hamiltonian $H_0$ is that of a system with only short range interactions. The presence of long range interactions in the post-quench dynamics of our interest is due to the form 
\begin{equation}
\Psi(\mathbf{x})=\int d\mathbf{r}\, \frac{\sigma(\mathbf{x}-\mathbf{r}/2) \sigma(\mathbf{x}+\mathbf{r}/2)}{|\mathbf{r}|^{\alpha}}\,,
\label{qo}
\end{equation}
of the quench operator, where $\sigma$ is an operator and $\alpha$ modulates the decay of these interactions. Working perturbatively in the quench size $\lambda$, we can perform our calculations in the basis of asymptotic states of the unperturbed theory, namely the short range interacting theory with Hamiltonian $H_0$. In this way we show that, for
\begin{equation}
\alpha>d\,,
\end{equation}
the one-point function of a local observable $\Phi$ behaves at large times as
\begin{equation}
\langle\Phi(t)\rangle=A_{\Phi,\sigma}+\lambda\bigg\{\sum_a \textrm{Re}\bigg[h^{\Phi,\sigma}_a\,e^{-iM_at}\bigg]+O(t^{-d/2})\bigg\}+O(\lambda^2)\,,
\label{result}
\end{equation}
where Re denotes the real part, $M_a>0$ is the mass of the quasiparticles of species $a$ of the pre-quench theory, and the constants $A_{\Phi,\sigma}$ and $h^{\Phi,\sigma}_a$ depend on $\alpha$ in a way that we determine. The result (\ref{result}) shows that at order $\lambda$ there are undamped oscillations unless the amplitudes $h^{\Phi,\sigma}_a$ vanish due to symmetry conditions that we also specify. We then explain that, as for the short range case \cite{oscill}, this result leads to undamped oscillations also when the higher orders in $\lambda$ are considered. This explains, in particular, why the undamped oscillations were numerically observed in \cite{Liu} in absence of small parameters. 

The paper is organized as follows. The next section is devoted to the determination of the post-quench state, while one-point functions are considered in section~\ref{1point} and their large time behavior in section~\ref{large_time}. Section~\ref{symmetries} illustrates the role of internal symmetries through the example of the Ising model, and the last section collects some final remarks.

\section{Post-quench state}
For negative times the system is in the ground state $|0\rangle$ of the pre-quench Hamiltonian $H_0$. The short range interacting pre-quench theory possesses a well defined basis of asymptotic quasiparticle states $|\mathbf{p}_1,...,\mathbf{p}_n\rangle$, where for the sake of notational simplicity we refer for the time being to the case of a single quasiparticle species; generalizations to more species are straightforward and will be considered when relevant. Excitations above the ground state correspond to quasiparticles modes with momentum $\mathbf{p}$ and energy $E_{\mathbf{p}}=\sqrt{M^2+\mathbf{p}^2}$, with $M>0$ the quasiparticle mass measuring the distance from a quantum critical point. 

The quench produces the nonequilibrium state
\begin{equation}
|\psi_0\rangle = S_{\lambda} |0\rangle = T\, \exp\left(-i \lambda \int_0^{\infty} dt \int d\mathbf{x}\, \Psi(\mathbf{x},t) \right) |0\rangle\,,
\end{equation}
where $T$ denotes chronological order and $S_{\lambda}$ is the operator whose matrix elements $\langle \mathbf{p}_1,...,\mathbf{p}_n|S_{\lambda}|0\rangle$ give the transition amplitude from the state $|0\rangle$ to the state $|\mathbf{p}_1,...,\mathbf{p}_n\rangle$. To first order in the quench size $\lambda$ we have\footnote{As usual, an infinitesimal
imaginary part is given to the energy to perform the time integral.}
\begin{align}
|\psi_0\rangle&\simeq |0\rangle -i\lambda \int_0^{\infty} dt \int d\mathbf{x} \int d\mathbf{r}\,\SumInt\limits_{n,\mathbf{p}_i} \SumInt\limits_{m,\mathbf{q}_j} \frac{1}{|\mathbf{r}|^{\alpha}}\, e^{i\left[\mathbf{P}\cdot \left(\mathbf{x}-\frac{\mathbf{r}}{2}\right)+\mathbf{Q}\cdot \mathbf{r}+Et \right]}\nonumber\\
&\times  |n\rangle\langle n|\sigma(0,0)|m\rangle\, \langle m|\sigma(0,0)|0\rangle \nonumber\\
& = |0\rangle +\lambda (2\pi)^d \int d\mathbf{r}\,\SumInt\limits_{n,\mathbf{p}_i} \SumInt\limits_{m,\mathbf{q}_j} \frac{\delta(\mathbf{P})}{E}\,\frac{1}{|\mathbf{r}|^{\alpha}}\, e^{i\mathbf{Q}\cdot \mathbf{r}}\, (F^{\sigma}_{m,n})^* F^{\sigma}_{m,0}\, |n\rangle\,,
\label{psi0_first}
\end{align}
where we introduced the compact notations\footnote{In the first sum in (\ref{sums}) we omit the term with $n=0$, which has $E=0$ and diverges in (\ref{psi0_first}) due to the energy denominator. This divergence, corresponding to vacuum renormalization, can be subtracted through a counterterm in the Hamiltonian \cite{quench}. In any case, this $n=0$ term would automatically cancel in the expression (\ref{onepoint}) for the one-point functions.}
\begin{equation} 
|n\rangle= |\mathbf{p}_1,...,\mathbf{p}_n\rangle\,,\hspace{2cm} |m\rangle = |\mathbf{q}_1,...,\mathbf{q}_m\rangle\,,
\end{equation}
\begin{equation}
\SumInt\limits_{n,\mathbf{p}_i} = \sum_{n=1}^{\infty} \frac{1}{n!} \int_{-\infty}^{\infty} \prod_{i=1}^n \frac{d\mathbf{p}_i}{(2\pi)^d E_{\mathbf{p}_i}}\,,\hspace{1.5cm}
\SumInt\limits_{m,\mathbf{q}_j} = \sum_{m=0}^{\infty} \frac{1}{m!} \int_{-\infty}^{\infty} \prod_{i=1}^m \frac{d\mathbf{q}_j}{(2\pi)^d E_{\mathbf{q}_j}}\,,
\label{sums}
\end{equation}
\begin{equation}
E=\sum_{i=1}^n E_{\mathbf{p}_i}\,, \quad \mathbf{P}=\sum_{i=1}^n \mathbf{p}_i\,,\hspace{1.5cm}
\tilde{E}=\sum_{j=1}^m E_{\mathbf{q}_j}\,,\quad \mathbf{Q}=\sum_{j=1}^m \mathbf{q}_j \,,
\end{equation}
as well as the notation
\begin{equation}
F_{m,n}^{\mathcal{O}}(\mathbf{q}_1,...,\mathbf{q}_m|\mathbf{p}_1,...,\mathbf{p}_n)=\langle \mathbf{q}_1,...,\mathbf{q}_m|\mathcal{O}(0,0)|\mathbf{p}_1,...,\mathbf{p}_n\rangle\,
\label{matrix_elements}
\end{equation}
for the matrix elements of local operators on the asymptotic states. Calling $\mathcal{P}$ the momentum operator, we also used the relation
\begin{equation}
\mathcal{O}(\mathbf{x}, t)= e^{i \mathcal{P}\cdot \mathbf{x}+i H_0 t}\, \mathcal{O}(0,0)\, e^{ -i\mathcal{P}\cdot \mathbf{x}-i H_0 t}\,.
\end{equation}

Defining $\mathbf{s}=|\mathbf{Q}|\mathbf{r}$, for $\alpha>d$ the integral over $\mathbf{s}$ converges at infinity and is regularized in the origin introducing a cutoff $|{\bf{s}}_0|\ll 1$. The result of this integral is then
\begin{equation}
|\psi_0\rangle = |0\rangle + \lambda D\SumInt\limits_{n,\mathbf{p}_i} \SumInt\limits_{m,\mathbf{q}_j} \frac{\delta(\mathbf{P})}{E}\, |\mathbf{Q}|^{\alpha-d}\, (F^{\sigma}_{m,n})^* F^{\sigma}_{m,0}\, |n\rangle +O(\lambda^2)\,,
\label{psi0}
\end{equation}
where $D$ is a constant depending on $\alpha$, $d$ and $|{\bf{s}}_0|$; the cutoff dependence makes $D$ nonuniversal, i.e. dependent on the microscopic details of the system.

The states $|\mathbf{q}_1,...,\mathbf{q}_m\rangle$ are those that we insert between the two $\sigma$ operators in (\ref{qo}) in order to resolve the bilocal product. For short wavelengths (i.e. for large momenta ${\bf q}_j$) these quasiparticles are sensitive to the nonuniversal microscopic details of the short-${\bf r}$ limit, and this is taken into account performing the ${\bf q}$-integrations for $|{\bf q}_j|<|{\bf q}_{\textrm{max}}|$. The presence of the cutoff ${\bf q}_{\textrm{max}}$ will be understood in the following.

\section{One-point functions}
\label{1point}
The variation of the one-point function of a local hermitian observable $\Phi({\bf{x}},t)$ with respect to its pre-quench value is given at first order in $\lambda$ by\footnote{The state normalization is $\langle 0|0\rangle=1$ at $t=-\infty$ and is preserved thereafter by the unitary time evolution.}
\begin{align}
\delta\langle \Phi(t)\rangle &= \langle \psi_0|\Phi(\mathbf{x},t)|\psi_0\rangle - \langle 0|\Phi(0,0)|0\rangle+C_{\Phi}\nonumber\\
&\simeq 2\lambda D \SumInt\limits_{n,\mathbf{p}_i} \SumInt\limits_{m,\mathbf{q}_j} \frac{\delta(\mathbf{P})}{E}\, |\mathbf{Q}|^{\alpha-d}\, \text{Re}\left\{(F^{\sigma}_{m,n})^* F^{\sigma}_{m,0} F^{\Phi}_{0,n}\, e^{-iEt} \right\} +C_{\Phi}\,,
\label{onepoint}
\end{align}
where the term
\begin{equation}
C_{\Phi}= -2\lambda D \SumInt\limits_{n,\mathbf{p}_i} \SumInt\limits_{m,\mathbf{q}_j} \frac{\delta(\mathbf{P})}{E}\, |\mathbf{Q}|^{\alpha-d}\, \text{Re}\left\{(F^{\sigma}_{m,n})^* F^{\sigma}_{m,0} F^{\Phi}_{0,n} \right\} 
\end{equation}
ensures continuity at $t=0$, namely the requirement $\delta \langle \Phi(0)\rangle=0$. The matrix elements (\ref{matrix_elements}) can be decomposed into the sum of a connected term
\begin{equation}
F_{m,n}^{\mathcal{O},c}(\mathbf{q}_1,...,\mathbf{q}_m|\mathbf{p}_1,...,\mathbf{p}_n)=\langle \mathbf{q}_1,...,\mathbf{q}_m|\mathcal{O}(0,0)|\mathbf{p}_1,...,\mathbf{p}_n\rangle_c\,
\end{equation}
plus disconnected terms\footnote{See \cite{lightcone} for the important role that disconnected parts play in correlation spreading out of equilibrium.} with delta functions associated to the annihilations of quasiparticles on the left with quasiparticles on the right. Explicitly,
\begin{align}
&F_{m,n}^{\mathcal{O}}(\mathbf{q}_1,...,\mathbf{q}_m|\mathbf{p}_1,...,\mathbf{p}_n)= F^{\mathcal{O},c}_{m,n}(\mathbf{q}_1,...,\mathbf{q}_m|\mathbf{p}_1,...,\mathbf{p}_n)\nonumber\\
&+\sum_{i=1}^n \sum_{j=1}^m (2\pi)^d E_{\mathbf{p}_i} \delta(\mathbf{p}_i-\mathbf{q}_j) F^{\mathcal{O},c}_{m-1,n-1}(\mathbf{q}_1,...,\mathbf{q}_{j-1},\mathbf{q}_{j+1},...,\mathbf{q}_m|\mathbf{p}_1,...,\mathbf{p}_{i-1},\mathbf{p}_{i+1},...,\mathbf{p}_n)\nonumber\\
&+ \sum\limits_{\substack{i,l=1\\i\neq l}}^n \sum\limits_{\substack{j,k=1\\j\neq k}}^m (2\pi)^{2d} E_{\mathbf{p}_i} E_{\mathbf{p}_l} \delta(\mathbf{p}_i-\mathbf{q}_j) \delta(\mathbf{p}_l-\mathbf{q}_k)\nonumber\\
&\times F^{\mathcal{O},c}_{m-2,n-2}(\mathbf{q}_1,...,\mathbf{q}_{j-1},\mathbf{q}_{j+1},...,\mathbf{q}_{k-1}, \mathbf{q}_{k+1},...,\mathbf{q}_m|\mathbf{p}_1,...,\mathbf{p}_{i-1},\mathbf{p}_{i+1},...,\mathbf{p}_{l-1}, \mathbf{p}_{l+1},...,\mathbf{p}_n)\nonumber\\
&+\cdots \,,
\label{decomposition}
\end{align}
where the dots indicate the terms with more than two annihilations. As a consequence of (\ref{decomposition}), the expectation value (\ref{onepoint}) can be expressed in terms of the connected matrix elements as
\begin{align}
\delta\langle \Phi(t)\rangle &\simeq 2\lambda D \SumInt\limits_{n,\mathbf{p}_i} \frac{\delta(\mathbf{P})}{E} \sum\limits_{m=0}^{\infty} \frac{1}{m!} \sum_{k=0}^{\text{min}(m,n)} \binom{n}{k} \frac{m!}{(m-k)!} \int_{-\infty}^{\infty}\prod_{j=1}^{m-k} \frac{d\mathbf{q}_j}{(2\pi)^d E_{\mathbf{q}_j}}\nonumber\\
&\times |\mathbf{q}_1+...+\mathbf{q}_{m-k}+\mathbf{p}_1+...+\mathbf{p}_k|^{\alpha-d}\, \text{Re}\big\{F^{\sigma,c}_{n-k,m-k}(\mathbf{p}_{k+1},...,\mathbf{p}_n|\mathbf{q}_1,...,\mathbf{q}_{m-k})\nonumber\\
&\times F^{\sigma,c}_{m,0}(\mathbf{q}_1,...,\mathbf{q}_{m-k},\mathbf{p}_1,...,\mathbf{p}_k|) F^{\Phi,c}_{0,n}(|\mathbf{p}_1,...,\mathbf{p}_n)\, e^{-iEt} \big\} +C_{\Phi}\,,
\label{onepoint_connected}
\end{align}
where we used the relation
\begin{equation}
[F_{m,n}^{\mathcal{O},c}(\mathbf{q}_1,...,\mathbf{q}_m|\mathbf{p}_1,...,\mathbf{p}_n)]^*= F_{n,m}^{\mathcal{O},c}(\mathbf{p}_1,...,\mathbf{p}_n|\mathbf{q}_1,...,\mathbf{q}_m)\,
\end{equation}
for the hermitian operators we consider.

\section{Large time behavior}
\label{large_time}
For $t\to \infty$ the rapid oscillations of $e^{-iEt}$ in (\ref{onepoint_connected}) suppress the integrals over momenta unless the phase is stationary, namely unless the momenta $\{\mathbf{p}_i\}$ are small. Consequently, we can use the nonrelativistic expression $E_{\mathbf{p}_i}\simeq M + \mathbf{p}_i^2/2M$ for the energies and rescale all momenta $\mathbf{p}_i$ by a factor $\sqrt{t}$. Since the matrix elements in (\ref{onepoint_connected}) generically go to constants in the limit $\{\mathbf{p}_i\}\to 0$ \cite{Barton}, the counting of the powers of time yields the large time behavior of the one-point function in the form
\begin{equation}
\delta \langle \Phi(t)\rangle \simeq \lambda \sum_{n=1}^{\infty} \text{Re}\left\{\left[ A^{\sigma,\Phi}_{m,n} \left(\frac{1}{\sqrt{t}} \right)^{(n-1)d} +B^{\sigma,\Phi}_{m,n} \left(\frac{1}{\sqrt{t}} \right)^{(n-2)d+\alpha}\right] e^{-inMt} \right\}+C_\Phi\,,
\label{largetime}
\end{equation}
where $A^{\sigma,\Phi}_{m,n}$ and $B^{\sigma,\Phi}_{m,n}$ are constants which in general depend on $\alpha$. The two different $t$-dependent contributions to (\ref{largetime}) come from the exam of the three types of terms present in (\ref{onepoint_connected}) depending on the number $k$ of annihilations, namely
\begin{itemize}
\item $k=m=0$:\\
\begin{equation}
2\lambda D \SumInt\limits_{n,\mathbf{p}_i} \frac{\delta(\mathbf{P})}{E}\, \text{Re}\big\{F^{\sigma,c}_{n,0}(\mathbf{p}_{1},...,\mathbf{p}_n|) F^{\sigma,c}_{0,0} F^{\Phi,c}_{0,n}(|\mathbf{p}_1,...,\mathbf{p}_n)\, e^{-iEt} \big\}
\end{equation}
\item $k=m\neq 0$:\\
\begin{align}
&2\lambda D \SumInt\limits_{n,\mathbf{p}_i} \frac{\delta(\mathbf{P})}{E} \sum\limits_{m=1}^{\infty} \binom{n}{m}\, |\mathbf{p}_1+...+\mathbf{p}_m|^{\alpha-d}\nonumber\\
&\times \text{Re}\big\{F^{\sigma,c}_{n-m,0}(\mathbf{p}_{m+1},...,\mathbf{p}_n|) F^{\sigma,c}_{m,0}(\mathbf{p}_1,...,\mathbf{p}_m|) F^{\Phi,c}_{0,n}(|\mathbf{p}_1,...,\mathbf{p}_n)\, e^{-iEt} \big\} 
\end{align}
\item $k\neq m$:
\begin{align}
& 2\lambda D \SumInt\limits_{n,\mathbf{p}_i} \frac{\delta(\mathbf{P})}{E} \sum\limits_{m=0}^{\infty} \frac{1}{m!} \sum_{k=0}^{\text{min}(m,n)} \binom{n}{k} \frac{m!}{(m-k)!} \int\prod_{j=1}^{m-k} \frac{d\mathbf{q}_j}{(2\pi)^d E_{\mathbf{q}_j}}\nonumber\\
&\times |\mathbf{q}_1+...+\mathbf{q}_{m-k}+\mathbf{p}_1+...+\mathbf{p}_k|^{\alpha-d}\, \text{Re}\big\{F^{\sigma,c}_{n-k,m-k}(\mathbf{p}_{k+1},...,\mathbf{p}_n|\mathbf{q}_1,...,\mathbf{q}_{m-k})\nonumber\\
&\times F^{\sigma,c}_{m,0}(\mathbf{q}_1,...,\mathbf{q}_{m-k},\mathbf{p}_1,...,\mathbf{p}_k|) F^{\Phi,c}_{0,n}(|\mathbf{p}_1,...,\mathbf{p}_n)\, e^{-iEt} \big\}\,.
\end{align}
\end{itemize}
Only in the second case all the momenta contained in the factor $|\cdots|^{\alpha-d}$ are ${\bf{p}}_i$'s and, due to the rescaling we perform on these momenta, produce an $\alpha$-dependent power of time, namely the second contribution in (\ref{largetime}).

We see that the leading contribution to (\ref{largetime}) at large time comes from $n=1$ and behaves as the sum of a purely oscillatory part plus a power $t^{-(\alpha-d)/2}$ which is suppressed in time in the regime $\alpha >d$ we are considering. However, when computing explicitly this term $n=1$, we realize that the term $m=k=1$, which should be the damped one, actually vanishes because of the factor $\delta(\mathbf{p})\, |\mathbf{p}|^{\alpha-d}$ in the integrand. Then at first order in $\lambda$ we have
\begin{align}
&\delta\langle \Phi(t)\rangle \simeq \frac{2\lambda D}{M^2} F_1^{\Phi} \bigg\{ F_1^{\sigma} F_0^{\sigma} \cos(Mt) + \sum_{m=1}^{\infty} \frac{1}{m!} \int\prod_{j=1}^{m} \frac{d\mathbf{q}_j}{(2\pi)^d E_{\mathbf{q}_j}}\, |\mathbf{q}_1+...+\mathbf{q}_{m}|^{\alpha-d}\nonumber\\
&\times \text{Re}\left[F^{\sigma,c}_{1,m}(\mathbf{0}|\mathbf{q}_1,...,\mathbf{q}_m) F^{\sigma,c}_{m,0}(\mathbf{q}_1,...,\mathbf{q}_m|) e^{-iMt}\right]\nonumber\\
&+ \sum_{m=2}^{\infty} \frac{1}{(m-1)!} \int\prod_{j=1}^{m-1} \frac{d\mathbf{q}_j}{(2\pi)^d E_{\mathbf{q}_j}}\, |\mathbf{q}_1+...+\mathbf{q}_{m-1}|^{\alpha-d}\nonumber\\
&\times \text{Re}\left[F^{\sigma,c}_{0,m-1}(|\mathbf{q}_1,...,\mathbf{q}_{m-1}) F^{\sigma,c}_{m,0}(\mathbf{q}_1,...,\mathbf{q}_{m-1},\mathbf{0}|) e^{-iMt}\right]\bigg\}+ O(t^{-d/2})+C_\Phi\,,
\label{n1}
\end{align}
where for the scalar hermitian operators of our interest  $F^{\mathcal{O},c}_{0,1}=F^{\mathcal{O},c}_{1,0}\equiv F_1^{\mathcal{O}}$ and $F^{\mathcal{O},c}_{0,0}\equiv F_0^{\mathcal{O}}$ are real constants. The $n=1$ term then entirely consists of undamped oscillations, while the first subleading term comes from $n=2$ and is suppressed as $t^{-d/2}$. It can be observed that in the short range limit, which is reached as $\alpha\to \infty$, the overall term Re$[b\,e^{-iMt}]$ resulting from sums and integrations should have $b$ real in order to match the overall $\cos(Mt)$ leading large time behavior originally obtained in \cite{quench} and (for $d$ dimensions) in \cite{oscillD}. It is also worth stressing that the analytical result (\ref{n1}) originates from our ability to {\it determine} the post-quench state, and cannot be obtained within more phenomenological approaches (see e.g. \cite{Cevolani1,Cevolani2}). For fixed $\alpha$, (\ref{n1}) holds up to a timescale that goes to infinity as $\lambda$ decreases. For the analogue of (\ref{n1}) in the short range case \cite{quench,DV}, the agreement with numerical simulations has been exhibited in \cite{Jacopo} with particularly high accuracy. 

We see that the frequency of the oscillations in (\ref{n1}) is the quasiparticle mass. Our derivation can be easily generalized to the case of several quasiparticle species with masses $M_a$. This leads to one oscillation frequency for each mass and to the result (\ref{result}), where $A_{\Phi,\sigma}$ is the sum of $\langle 0|\Phi|0\rangle$ plus the multi-species generalization of $C_\Phi$.

The result (\ref{result}) shows that at order $\lambda$ there are undamped oscillations unless the amplitudes $h^{\Phi,\sigma}_a$ all vanish. We now explain that, when such vanishing does not occur, the first order result actually implies undamped oscillations in the full result (first order plus higher orders). Let us focus on the remainder of (\ref{result}), namely the resummation of all terms beyond first order, which we call $f(t)$. From the mathematical point of view, $f(t)$ is a function of time that for $t\to\infty$ can either diverge, or approach a constant value, or itself display undamped oscillations. However, the first possibility cannot occur for ordinary physical observables such as a local magnetization; indeed, at a given point in space this can grow in time at most to the limit of maximal ordering, but cannot diverge. Once discarded the possibility of divergence, the other two possibilities lead to undamped oscillations of the complete result (first order plus $f(t)$) as long as these are present at first order\footnote{Of course, this does not imply that the oscillations of the complete result quantitatively coincide with those of the first order.}. This observation does not depend on the range of the interactions and was originally made in \cite{oscill} for the short range case. 

The detailed nonperturbative origin of the undamped oscillations was exhibited in \cite{structure} for the short range case. Their physical meaning is clear and does not depend on the range of the interactions. For an isolated translation invariant system, the energy injected by the quench cannot be dissipated nor redistributed\footnote{By this we mean that after the quench the energy density is the same and remains constant at all points in space. This has to be contrasted with the case of quenches performed only in a finite region of the space $\mathbb{R}^d$ occupied by the system. In such a case the energy density introduced by the quench will go everywhere to zero as $t\to\infty$ and will be unable to sustain undamped oscillations \cite{oscill,oscillD}.} within the system. Hence, nothing can prevent some oscillations induced by the quench to remain undamped in some cases. We showed that nonperturbative undamped oscillations are present when at least one of the amplitudes $h^{\Phi,\sigma}_a$ in (\ref{result}) does not vanish. In the next section we illustrate how the vanishing or not of these amplitudes can be predicted on symmetry grounds.

We conclude this section with a comment on the role of $\alpha$ in our results. We have seen that $\alpha$ enters (\ref{result}) only through the amplitudes $h^{\Phi,\sigma}_a$, so that the large time behavior of one-point functions is not  qualitatively affected by variations of $\alpha>d$. This result can be understood observing that, morally, $\alpha$ determines a scale $R_\alpha$, which decreases as $\alpha$ increases, beyond which the long range interactions have decayed enough to become negligible. Hence, $\alpha$ can be expected to play an important role only when the nonequilibrium dynamics is studied on space-time scales which do not substantially exceed $R_\alpha$. In particular, $\alpha$ is not expected to play an important role for one-point functions when $t\to\infty$ for $\alpha$ fixed, which is the case we study. Our analytical results substantiate these heuristic reasonings. On similar grounds, we expect that replacing $r^{-\alpha}$ in (\ref{qo}) with a more general function $g_\alpha(r)$ decaying as $r^{-\alpha}$ for large $r$ should not qualitatively change the result (\ref{result}). Like the `ultraviolet' cutoffs ${\bf s}_0$ and ${\bf q}_{\textrm{max}}$ discussed in section~2, the behavior of $g_\alpha(r)$ on short/medium distances should only affect the amplitudes $h^{\Phi,\sigma}_a$.

\section{Role of symmetries: Ising model}
\label{symmetries}
In this section we illustrate the role that internal symmetries play for the results of the previous section through the basic example of the $d$-dimensional quantum Ising ferromagnet in a transverse field with long range exchange interactions. We then consider the post-quench Hamiltonian
\EQ
H_\textrm{Ising}=-J_0\sum_{\langle i,j\rangle}\sigma^x_i\sigma^x_j-\sum_{i,j}\frac{J}{r_{ij}^\alpha}\, \sigma^x_i\sigma^x_j-h\sum_i\sigma^z_i\,,
\label{Ising}
\EN
where $\sigma^{x,z}_i$ are Pauli matrices at site $i$ of a $d$-dimensional regular lattice, $\langle i,j\rangle$ denotes nearest neighbor interaction, $h$ is the transverse magnetic field, and $r_{ij}$ is the distance (in lattice units) between sites $i$ and $j$. The pre-quench theory corresponds to $J=0$ and for $|h|=h_c$ possesses a quantum critical point associated to the spontaneous breaking of spin reversal ($\mathbb{Z}_2$) symmetry in the $x$ direction. The operator $\sigma^z_i$ ($\sigma^x_i$) is $\mathbb{Z}_2$-even (odd). The paramagnetic (ferromagnetic) phase corresponds to $|h|>h_c$ ($|h|<h_c$). These phases persist for $J>0$ with a modified value of the critical field. 

In the notations of our previous derivations, the present case corresponds to $\lambda\sim J$ and a quench operator (\ref{qo}) specified by $\sigma=\sigma^x$. We further distinguish the case in which the quench is performed within the paramagnetic phase from that in which it is performed within the ferromagnetic phase. This distinction proceeds from symmetry considerations in the pre-quench theory. In the paramagnetic phase, the fundamental quasiparticle excitation is created by the order parameter operator $\sigma^x({\bf x})$ and is then $\mathbb{Z}_2$-odd. It follows that the matrix elements  $F_{m,n}^{\sigma^z}$ are nonzero only if $m+n$ is even, while the matrix elements $F_{m,n}^{\sigma^x}$ are nonzero only if $m+n$ is odd. Hence, after the quench, (\ref{onepoint_connected}) yields $\langle\sigma^x(t)\rangle=0$ and $\langle\sigma^z(t)\rangle\neq 0$, as a priori expected from the symmetry. On the other hand, (\ref{n1}) shows that $\langle\sigma^z(t)\rangle$ will go to a constant at large times, since the amplitudes of the undamped oscillations vanish by symmetry.

In the ferromagnetic phase, instead, the symmetry is spontaneously broken and the matrix elements $F_{m,n}^{\sigma^z}$ and $F_{m,n}^{\sigma^x}$ are all nonzero in $d>1$; the case $d=1$ will be separately discussed in a moment. Hence, for $d>1$, (\ref{n1}) shows that $\langle\sigma^x(t)\rangle$ and $\langle\sigma^z(t)\rangle$ both possess undamped oscillations after the quench within the ferromagnetic phase\footnote{It is worth stressing that we only rely on ordinary symmetries and work in any dimension. In particular, we do not invoke special symmetries related to integrability of some one-dimensional systems (see e.g. \cite{MBJ}).}. As we know from (\ref{result}), at order $\lambda$ the frequencies of the oscillations consistently are the masses of the quasiparticles of the pre-quench theory\footnote{For the ferromagnetic phase with short range interactions in $d=2$ there is numerical consensus \cite{CH,LSW,CHPZ,DKSTV,Nishiyama,RBLD} on the existence, in addition to the lightest quasiparticle with mass $M_1$, of a second quasiparticle with mass $M_2\approx 1.8 M_1$.}. The expectation is that, as analytically shown in \cite{structure} in the short range case, for larger quenches the oscillation frequencies will be determined by the masses of the post-quench Hamiltonian\footnote{Including frequencies equal to mass differences which in perturbation theory arise at order $\lambda^2$, see \cite{oscill,structure}.}. 

The quench within the ferromagnetic phase in $d=1$ requires a separate discussion because in the short range interacting pre-quench theory the excitations have a topological nature -- they are kinks, see \cite{review} for a review -- and couple to $\sigma^x$ and $\sigma^z$ only in topologically neutral pairs. It follows that $F_{m,n}^{\sigma^z}$ and $F_{m,n}^{\sigma^x}$ are nonzero only if $m+n$ is even, so that the amplitudes of the undamped oscillations vanish in (\ref{n1}). This is, however, a very specific case in which the perturbative result (\ref{n1}) does not directly apply due to an additional peculiarity of the one-dimensional Ising model. Indeed, while in spontaneously broken phases of one-dimensional theories with short range interactions kinks are ubiquitous\footnote{See \cite{q_int,CDGJM,AT_correlators,LTD,DV_interfaces} for physical effects due to the kinks of the Potts and Ashkin-Teller models in and out of equilibrium.}, in models other than Ising they interact, and normally the $O(\lambda)$ long range perturbation will slightly shift the masses without qualitatively changing the quasiparticle spectrum\footnote{This isospectrality for quenches within a same phase is generally true -- also for Ising -- in the short range limit $\alpha\to\infty$, since the quench only rescales the masses. The $O(\lambda$) mass shift contributes to the $O(\lambda^2)$ correction to one-point functions \cite{quench,DV}.}. For Ising, instead, the kinks do not interact in the pre-quench theory (see \cite{review}), and an arbitrarily small attractive interaction introduced by the quench with $\alpha<\infty$ produces bound states\footnote{It is known that in $d=1$ any attractive potential produces bound states \cite{Landau}.}. Since topologically neutral bound states can couple to $\sigma^x$ and $\sigma^z$ in any number, there is no symmetry left to cause the vanishing of the amplitudes of the undamped oscillations in (\ref{result}) for $\alpha<\infty$. These undamped oscillations have indeed been numerically observed in \cite{Liu} in a quench from the ground state within the ferromagnetic phase\footnote{Absence of relaxation of local observables to a constant value is a known feature of long range interacting systems with $\alpha<d$ \cite{Kastner,Defenu}, which was recently predicted to extend also to open systems \cite{MLC}. It would be interesting to see if some connection can be established with the asymptotic states formalism, which in the present paper has been used for $\alpha>d$.}.

The considerations we developed for Ising can be extended along similar lines to other models. The form (\ref{qo}) of the quench operator generally covers power law decaying interactions between two spin components. The features specific to each model will be determined by the internal symmetry and the quasiparticle spectrum of the pre-quench (short range) theory which, for example, are discussed in \cite{q_int} for the Potts and Ashkin-Teller spin chains, and in \cite{DV} for the XYZ chain.

\section{Conclusion}
In this paper we showed how the general asymptotic state theory of quantum quenches extends to the case of $d$-dimensional systems with long range interactions decaying as $r^{-\alpha}$. We succeeded in generally and analytically accessing this type of dynamics treating the long range interaction as switched on by the quench, the system being before that in the ground state of an Hamiltonian with short range interactions only. This allowed us to perform our derivations in the basis of asymptotic quasiparticle states of the pre-quench theory, which is endowed with all the properties holding for short range interacting systems. In this way we determined the state produced by the quench and encoding nonequilibrium behavior in presence of the long range interactions. We could then write down the expression for the one-point functions of local observables such as the order parameter and analyze their behavior at large times. These derivations were performed for arbitrarily strong interactions among the quasiparticles of the pre-quench theory and perturbatively in the quenche size $\lambda$. We further observed that characteristic features of the dynamics persist nonperturbatively. This is the case, in particular, for oscillations of the order parameter which remain undamped in time under conditions determined by the internal symmetries of the system. This result substantiates the intuition that in isolated homogeneous systems -- which cannot dissipate or internally redistribute their energy -- some oscillations induced by the quench can remain undamped at large times. The theory explains when this happens and yields frequencies and amplitudes.

\vspace{1cm}
{\bf Acknowledgements}. ND acknowledges funding by the Swiss National Science Foundation (SNSF) under project ID: 200021 207537, by the Deutsche Forschungsgemeinschaft (DFG, German Research Foundation) under Germany’s Excellence Strategy EXC2181/1-390900948 (the Heidelberg STRUCTURES Excellence Cluster) and by the European Union under GA No. 101077500–QLR-Net, and partial support by grant NSF PHY-230935 to the Kavli Institute for Theoretical Physics (KITP).

\end{document}